\documentclass[osajnl,twocolumn,showpacs,superscriptaddress,11pt,floatfix]{revtex4-1}

\usepackage[british]{babel}
\usepackage{amsmath}
\usepackage{amssymb}
\usepackage{graphicx}
\usepackage{xspace}

\newcommand{\linklength}{$2\times298$ km}
\newcommand{\sups}[1]{\ensuremath{^{\text{#1}}}\xspace}	
\newcommand{\pow}[1]{\ensuremath{\,\times\,10^{#1}}\xspace}
\newcommand{\td}{\ensuremath{\text{d}}\xspace} 
\newcommand{\anbrace}[1]{\ensuremath{\left({#1}\right)}\xspace}
\newcommand{\degrees}{\sups{$\circ$}{}}
\newcommand{\e}[1]{\ensuremath{e^{#1}}\xspace}		
\newcommand{\fo}{\ensuremath{f_{0}}\xspace}
\newcommand{\frep}{\ensuremath{f_{\text{rep}}}\xspace}
\newcommand{\fceo}{\ensuremath{f_{\text{CEO}}}\xspace}
\newcommand{\fbeat}{\ensuremath{f_{\text{beat}}}\xspace}

\begin{document}
\title{Effect of soil temperature on one-way optical frequency transfer through dense-wavelength-division-multiplexing fibre links}

\author{T. J. Pinkert}
\affiliation{Department of Physics and Astronomy, LaserLaB, VU University, De Boelelaan 1081, 1081 HV Amsterdam, The Netherlands}

\author{O. B\"oll}
\author{L. Willmann}
\affiliation{Van Swinderen Institute for Particle Physics and Gravity, Faculty of Mathematics and Natural Sciences, Nijenborgh 4, 9747 AG Groningen, The Netherlands}

\author{G. S. M. Jansen}
\affiliation{Department of Physics and Astronomy, LaserLaB, VU University, De Boelelaan 1081, 1081 HV Amsterdam, The Netherlands}

\author{E. A. Dijck}
\author{B. G. H. M. Groeneveld}
\affiliation{Van Swinderen Institute for Particle Physics and Gravity, Faculty of Mathematics and Natural Sciences, Nijenborgh 4, 9747 AG Groningen, The Netherlands}

\author{R. Smets}
\affiliation{SURFnet, Radboudkwartier 273, 3511 CK Utrecht, The Netherlands}

\author{F. C. Bosveld}
\affiliation{KNMI, Utrechtseweg 297, 3731 GA De Bilt, The Netherlands}

\author{W. Ubachs}
\affiliation{Department of Physics and Astronomy, LaserLaB, VU University, De Boelelaan 1081, 1081 HV Amsterdam, The Netherlands}

\author{K. Jungmann}
\affiliation{Van Swinderen Institute for Particle Physics and Gravity, Faculty of Mathematics and Natural Sciences, Nijenborgh 4, 9747 AG Groningen, The Netherlands}

\author{K. S. E. Eikema}
\author{J. C. J. Koelemeij}\email{Corresponding author: j.c.j.koelemeij@vu.nl}
\affiliation{Department of Physics and Astronomy, LaserLaB, VU University, De Boelelaan 1081, 1081 HV Amsterdam, The Netherlands}

\begin{abstract}%
Results of optical frequency transfer over a carrier-grade dense-wavelength-division-multiplexing (DWDM) optical fibre network are presented. The relation between soil temperature changes on a buried optical fibre and frequency changes of an optical carrier through the fibre is modelled. Soil temperatures, measured at various depths by the Royal Netherlands Meteorology Institute (KNMI) are compared with observed frequency variations through this model. A comparison of a nine-day record of optical frequency measurements through the \linklength{} fibre link with soil temperature data shows qualitative agreement. A soil temperature model is used to predict the link stability over longer periods (days-months-years). We show that one-way optical frequency dissemination is sufficiently stable to distribute and compare e.g. rubidium frequency standards over standard DWDM optical fibre networks.
\end{abstract}


\maketitle

\section{Introduction}
In recent years, fibre-optic connections in telecommunication networks have proven to be suitable for frequency comparisons and frequency distribution with high stability over long distances. In general the frequency of either an ultra-stable continuous-wave (CW) laser or a stable microwave reference is transmitted over an optical fibre and received at the remote site. Several experiments~\cite{lit:pap:JOSAB-williams-2008,lit:pap:EPJD-lopez-2008,lit:proc:jaldehag-2009,lit:pap:OL-grosche-2009,lit:pap:OE-fujieda-2011,lit:pap:S-predehl-2012,lit:pap:OE-lopez-2012} have shown that frequency comparisons at or below the current accuracy level of the best atomic frequency references, a few times 10\sups{-18}~\cite{lit:pap:S-rosenband-2008,lit:pap:PRL-chou-2010,lit:pap:S-hinkley-2013,lit:pap:N-bloom-2014}, are feasible over long-haul fibre connections. For example, the optical frequency of the 1S-2S transition in atomic hydrogen was recently measured with respect to a remote Cs frequency standard with 4.5\pow{-15} relative uncertainty, employing a 920 km long fibre-optic link performing at the $4\pow{-19}$ uncertainty level~\cite{lit:pap:S-predehl-2012,lit:pap:PRL-matveev-2013}.

Fibre-optic methods for remote frequency comparison can provide significantly better stability than current satellite-based methods. These include two-way satellite time and frequency transfer (TWSTFT)~\cite{lit:proc:hanson-1989,lit:proc:zhang-2009,lit:pap:M-piester-2008} and (carrier-phase) common-view global positioning system ((CP/)CV-GPS) comparisons~\cite{lit:proc:allan-1980,lit:proc:zhang-2000,lit:pap:CLMIJM-lombardi-2001}. The accuracy limits of satellite based methods are on the order of $10^{-15}$ at one day~\cite{lit:pap:ITUFFC-larson-1999,lit:proc:brown-2000,lit:pap:CLMIJM-lombardi-2001}, which is already insufficient to compare state-of-the-art cesium fountain clocks operating at less than $5\pow{-16}$ uncertainty level~\cite{lit:pap:M-weyers-2011,lit:pap:M-heavner-2014}.

\begin{figure}[htb]
    \centering
    \includegraphics[width=\columnwidth]{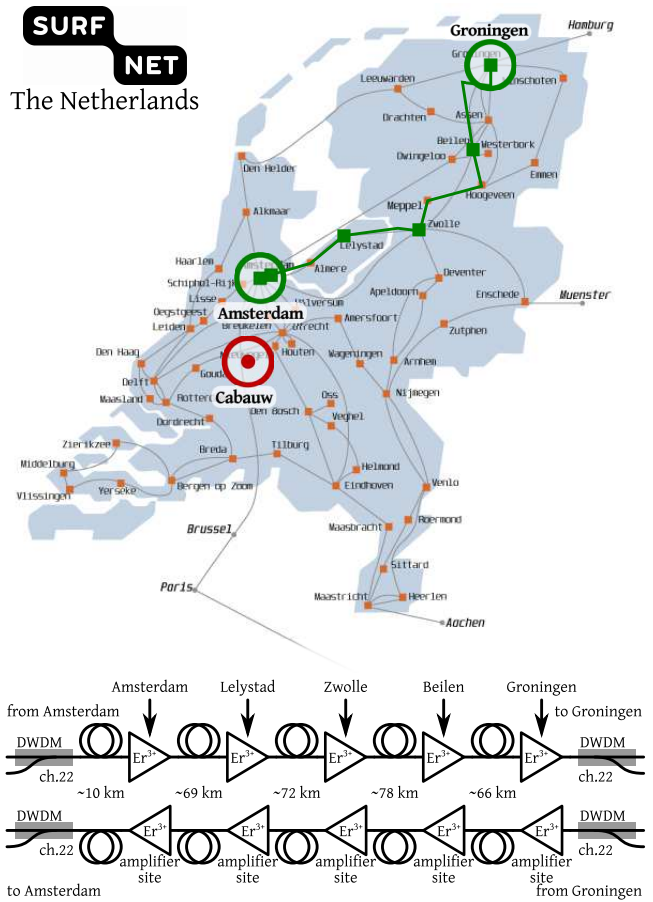}
    \caption{Top: Schematic map of the SURFnet fibre-optic network. (green line) The fibre link between VU University Amsterdam and Van Swinderen Institute Groningen. (green squares) Amplifier sites. (red circle) The KNMI measurement site at Cabauw. Bottom: Schematic representation of the duplex fibre link between Amsterdam and Groningen in the SURFnet network (length: $2\times295$ km). The Er\sups{3+}-amplifiers are used by all active DWDM channels. In Amsterdam $\sim500$ m of intra-office fibre bridges the distance between the SURFnet node and the laboratory. In Groningen $\sim2$ km of additional underground fibre is needed to bridge the distance from the SURFnet node at the computing centre of the University of Groningen to the laboratory, adding to a total link length of \linklength{}.}\label{fig:surfnet_fibre_network}
\end{figure}

Plans for high-resolution laser spectroscopy experiments at VU University Amsterdam LaserLaB and at Van Swinderen Institute, University of Groningen,  would be greatly facilitated by direct frequency comparisons over an optical fibre-link at stability levels better than ${10}^{-14}$. For this purpose, a \linklength{} fibre-optic connection between both laboratories has been established using a $2\times295$ km DWDM channel provided by SURFnet. The top part of Fig.~\ref{fig:surfnet_fibre_network} gives an overview of the SURFnet fibre network in the Netherlands. The optical path between VU University Amsterdam and Van Swinderen Institute Groningen is marked in green. The bottom part shows the details of the optical path.

The link consists of two unidirectional fibres. Interferometric detection and active compensation of fibre length changes~\cite{lit:pap:OL-ma-1994} is therefore not possible. However, it is possible to create a bidirectional path in the optical fibre, and implement a compensation system~\cite{lit:pap:OE-lopez-2012}. It must be noted that, in contrast to some of the other frequency comparison links~\cite{lit:pap:JOSAB-williams-2008,lit:pap:EPJD-lopez-2008,lit:pap:OL-grosche-2009,lit:pap:OE-fujieda-2011,lit:pap:S-predehl-2012}, our link is part of a carrier-grade DWDM optical network in which the fibre is shared with other users, and several other wavelength channels are simultaneously used for data transfer.

The use of a standard DWDM channel enables us to characterise the performance of unidirectional transmission of optical frequency references in public transport networks carrying live network traffic. It also allows a systematic study of environmental conditions on the link stability, such as soil temperature variations, which is the main aim of this work. Unidirectional frequency transfer will be useful for institutions and industries which require accurate and reliable time and frequency references, e.g. for calibration of equipment. For example, optical frequency distribution may be used for accurate length and, in the future, mass measurements (through the Josephson effect and the Watt balance) by referencing to the SI second.

This article is structured as follows. In Sec.~\ref{sec:analysis} a model for the influence of temperature on the stability of fibre-optic frequency transfer is presented, along with a soil temperature model. In Sec.~\ref{sec:method}, we present our approach to determine the stability of the \linklength{} optical frequency link, as well as details of the setups used in Amsterdam and Groningen. Results are discussed in Sec.~\ref{sec:measurement}, followed by conclusions and an outlook presented in Sec.~\ref{sec:conclusions}.

\section{Theory: frequency stability of the fibre link and soil temperature}\label{sec:analysis}
The phase $\varphi$ accumulated by a monochromatic light wave guided along a certain path of length $L$ and effective refractive index $n$ can be written as
\begin{equation}\label{eq:link_length_in_phase}
    \varphi = \frac{\omega_{0}}{c}nL,
\end{equation}
where $\omega_{0}$ is the frequency of the light (in radians per second). Several physical processes can lead to phase (and thus frequency) variations in fibre-optic links.

At short time scales ($<100$ sec) environmental vibrations couple to the fibre and can therefore cause path length variations, e.g. via stress-induced refractive index variation, which may occur at frequencies up to tens of kHz. At time scales longer than 1 s, significant phase variations also occur because of thermal expansion of the fibre and thermally induced changes in the refractive index. These variations are typically slow and, for a fibre which is installed mainly underground in an outdoor environment, they are mostly affected by the diurnal and seasonal soil temperature cycles~\cite{lit:pap:SSSAJ-elias-2004}. Our work is focused on the understanding of the long-term stability of fibre links in relation to such temperature variations.

The phase variations due to a time-varying temperature $T$ are
\begin{equation}\label{eq:phase_variation_due_to_temperature}
    \frac{\td\varphi}{\td t} = \frac{\omega_{0}}{c}\anbrace{L\frac{\partial n}{\partial T}\frac{\td T}{\td t} + n\frac{\partial L}{\partial T}\frac{\td T}{\td t}}.
\end{equation}
We can express the relative length variations of the fibre as a function of temperature as
\begin{equation}
    \frac{1}{L}\frac{\partial L}{\partial T}=\alpha_{\Lambda},
\end{equation}
where $\alpha_{\Lambda}$ is the thermal expansion coefficient of the fibre. Furthermore it is customary to write
\begin{equation}
    \frac{\partial n}{\partial T}=\alpha_{n},
\end{equation}
with $\alpha_{n}$ the thermo-optic coefficient. Note that both $\alpha_{\Lambda}$ and $\alpha_{n}$ are weakly dependent on temperature, which we ignore here.

A typical (room-temperature) value of the thermal expansion coefficient of the fibre glass is $\alpha_{\Lambda}=5.6\pow{-7}$/ \degrees C~\cite{lit:pap:IPO-lin-2004}. The thermo-optic coefficient has a typical value of $\alpha_{n}=1.06\pow{-5}$/\degrees C~\cite{lit:pap:IPO-lin-2004}, and is therefore the main cause of phase variations due to temperature changes. Similar effects due to varying air pressure are approximately two orders of magnitude smaller~\cite{lit:pap:JLT-ghosh-1998} and have therefore not been included in the model.

Considering the heat flux in isotropic media (soil) for a vertical temperature gradient and varying temperature, and modelling the temperature variation as a sinusoidal periodic signal, Van der \nobreak{Hoeven} and \nobreak{Lablans}~\cite{lit:rep:hoeven-1992} derive the equation for the temperature of the soil at a certain depth $z$ and time $t$ as
\begin{multline}\label{eq:soil_temperature_hoeven}
    T(z,t)=T_{0} + A_{T_{0}}\e{-zC_{\varphi}}\\
    \times\sin\anbrace{\frac{2\pi}{P_{T_{0}}}\anbrace{t-t_{0}} - zC_{\varphi}},
\end{multline}
where $T_{0}$ is the average temperature at the surface ($z=0$), $A_{T_{0}}$ is the amplitude of the temperature variation at the surface with period $P_{T_{0}}$, and $t_{0}$ is an arbitrary time offset. The phase constant
\begin{equation}\label{eq:phase_constant}
    C_{\varphi}=\frac{1}{C_{s}}\sqrt{\frac{\pi}{P_{T_{0}}}},
\end{equation}
includes the soil constant $C_{s}$ which is determined by the thermal conductivity $\lambda$, the specific heat capacity $C_{m}$, and the mass density $\rho$ of the soil according to
\begin{equation}\label{eq:soil_constant}
    C_{s}=\sqrt{\lambda/\rho C_{m}}.
\end{equation}
Equation~\eqref{eq:soil_temperature_hoeven} can be applied to both diurnal and annual variations in temperature. A remark must be made that accurate modelling of the soil temperature is delicate and involves, among others, the groundwater levels and groundwater freezing rates in winter~\cite{lit:rep:hoeven-1992}.

In this model the amplitude of the temperature wave decreases exponentially with depth, while it undergoes a phase shift linear with depth. The equation is universal and can also be applied to other materials as long as the physical properties $\lambda$, $\rho$ and $C_{m}$ are known.

With the help of Eq.~\eqref{eq:soil_temperature_hoeven}, average frequency deviations $\Delta f$ from the nominal frequency $\fo=\omega_{0}/2\pi$ can be calculated for any depth, using Eq.~\eqref{eq:phase_variation_due_to_temperature} for temperature differences $\Delta T$ over a time interval $\Delta t$ as
\begin{equation}\label{eq:frequency_from_temperature}
        \Delta f = 2\pi f_{0}\frac{L}{c}\anbrace{\alpha_{n}+n\alpha_{\Lambda}}\frac{\Delta T}{\Delta t}.
\end{equation}

\section{Experimental methods}\label{sec:method}
In order to characterise the frequency stability of the fibre link, two different methods are used. The first method consists of the one-way transmission of a C-band-wavelength CW laser, locked to a mode of an Er\sups{3+}-doped fibre frequency comb laser in Amsterdam. In Groningen, the transmitted laser frequency is measured using a similar frequency comb laser. Both frequency combs are locked to GPS-disciplined atomic clocks. Therefore, the stability of the optical frequency measurement in Amsterdam and in Groningen depends on the fibre link as well as the atomic clock stabilities. The second method employs a closed fibre loop Amsterdam -- Groningen -- Amsterdam. In this case the laser frequency can be compared with itself after its roundtrip through the fibre loop. The second method takes advantage of the fact that the laser frequency instability on the time scale of the measurement is much smaller than the instabilities introduced in the fibre loop.

\begin{figure*}[!htb]
    \centering
    \includegraphics[width=0.714\textwidth]{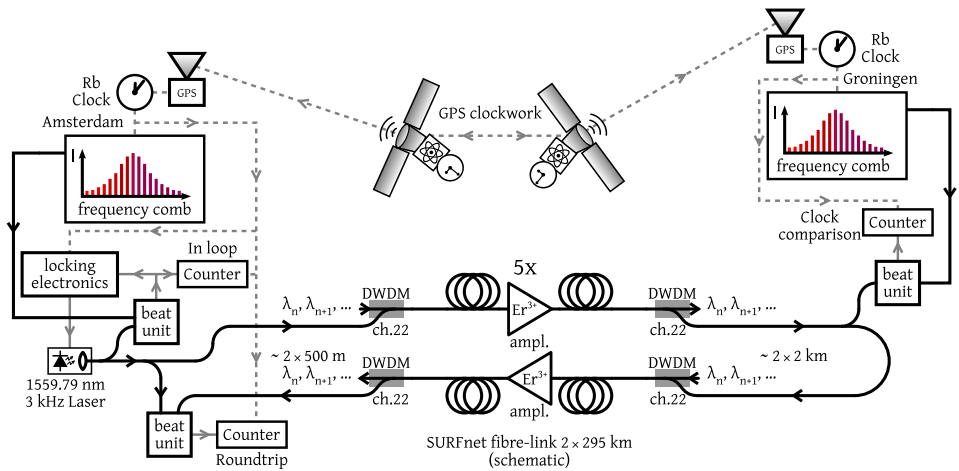}
    \caption{The Amsterdam -- Groningen fibre link, overview of the experimental setup. This arrangement allows for optical-versus-GPS comparisons (essentially an fibre-optic frequency comparison of the GPS-linked Rb clocks in Amsterdam and Groningen), and for measurements of the roundtrip stability of the fibre-link in Amsterdam. Details of the CW laser lock setup are given in Fig.~\ref{fig:fibre_network_experimental_setup_laser_locking}. The setup for the roundtrip analysis and for the optical versus Rb/GPS comparison are shown in detail in Fig.~\ref{fig:fibre_network_experimental_setup_roundtrip_measurement} and Fig.~\ref{fig:fibre_network_experimental_setup_clock_comparison}, respectively.}\label{fig:fibre_network_experimental_setup}
\end{figure*}

Figure~\ref{fig:fibre_network_experimental_setup} gives an overview of the measurement setup that is used to characterise the frequency stability of the fibre link. A narrow-linewidth CW laser (Redfern Integrated Optics inc. (RIO) Planex, 3 kHz linewidth, 20 mW output power, and wavelength 1559.79 nm [ITU channel 22]) is phase-locked to the fibre frequency comb laser (Menlo Systems FC1500) and used as an absolute optical frequency reference. To verify the lock quality, the in-loop beat signal is counted. The optical reference frequency is sent to Groningen via the fibre link (injected power 0 dBm). The nominal length $L$ of the fibre-link is \linklength{} between the laboratories in Amsterdam and Groningen. The optical fibres of the pair are located in a fibre-bundle and thus follow nearly the same physical path.

In Groningen the light is split. Part of the light is used in a frequency comparison against the local optical frequency standard (Menlo Systems FC1500), while the other part of the light is sent back to Amsterdam (received power 0--6 dBm). As pointed out above, the comparison of the frequency of the light after the roundtrip with the CW laser source reveals the noise contribution from the fibre link.

Both optical frequency combs are locked to rubidium (Rb) frequency standards (SRS FS725), which are disciplined to the 1 pps output of GPS receivers (Amsterdam: Trimble Acutime 2000, Groningen: Navsync CW46). The combined instability of the GPS-receiver output and Rb clocks is transferred to the frequency comb lasers via the various rf locks used to stabilise the frequency comb laser repetition rate frequency \frep and carrier envelope offset frequency \fceo.

\begin{figure}[htb]
    \centering
    \includegraphics[width=\columnwidth]{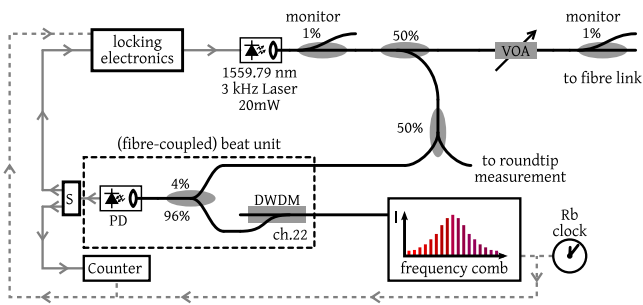}
    \caption{CW laser stabilisation setup (Amsterdam). The 1559.79 nm, 3 kHz diode laser is frequency stabilised by a phase-locked loop to a mode of the Er\sups{3+}-fibre frequency comb laser. The photo-diode signal of the fibre-coupled beat unit is amplified, filtered and split by a 3 dB power splitter (S) for input to the phase detector and the counter. The stabilisation setup is fully referenced to the GPS-disciplined Rb frequency standard. The monitor ports are used to observe optical power variations of the Planex laser before and after the variable optical attenuator (VOA), which regulates the laser power to a constant level before injection into the telecommunication network.}\label{fig:fibre_network_experimental_setup_laser_locking}
\end{figure}

Figure~\ref{fig:fibre_network_experimental_setup_laser_locking} shows the details of the CW laser stabilisation setup. Part of the light of the diode is split off and fed to a fibre-based beat-note unit consisting of a DWDM filter to reject a large part of the frequency comb spectrum, a fused coupler, and a fibre-coupled photodiode ($\sim2$ GHz bandwidth, 50/125 $\mu$m multi-mode fibre coupled) to detect the rf beat signal. The rf beat signal ($\fbeat=60$ MHz) is bandpass filtered (filter bandwidth $>10$ MHz) and amplified before comparison with a signal generator (Agilent 33250A, referenced to the Rb clock), via counting phase detector electronics. Feedback on the diode laser phase is achieved via a fast PID controller acting directly onto the diode-laser injection current (bandwidth $>1$ MHz).

To verify proper CW-laser locking conditions during the experiment, the in-loop rf beat-frequency is recorded with a Rb-referenced counter. For dead-time-free counting of rf frequencies, either a quasi-continuous double Agilent 53132A counter setup is used, or an Agilent 53230A counter in a continuous reciprocal frequency counting mode. Finally, a regulated variable optical attenuator ensures that the optical power injected into the fibre link remains constant.

\begin{figure}[htb]
    \centering
    \includegraphics[width=\columnwidth]{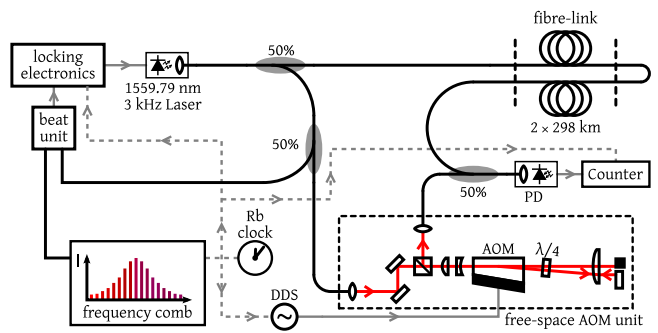}
    \caption{Experimental setup for the characterisation of the passive frequency stability of the fibre link (Amsterdam). For the long roundtrip measurements the free-space AOM unit (300 MHz) was replaced by a fibre-coupled AOM ($-42$ MHz). In both cases the AOM was driven by a Rb-referenced DDS unit whith a set accuracy of $\sim3.55$ $\mu$Hz. Frequency deviations of the link are recorded with a Rb-referenced counter.}\label{fig:fibre_network_experimental_setup_roundtrip_measurement}
\end{figure}

To characterise the stability of the fibre link, the setup of Fig.~\ref{fig:fibre_network_experimental_setup_roundtrip_measurement} is used. Light of the diode laser is split by a fused coupler. Part of the light is sent via the roundtrip Amsterdam -- Groningen -- Amsterdam, while another portion of the light is frequency shifted by an acousto-optic modulator (AOM). The output of the AOM is combined with the roundtrip optical signal after transmission by the fibre link (power $\sim-3$ dBm) and the beat frequency is detected with a fibre-coupled photo-diode. Any frequency variations introduced by the fibre link can be measured as frequency deviations from the AOM frequency, which is generated by a Rb-referenced DDS (Analog Devices AD9912).

To measure the roundtrip stability two AOM + counter setups were used in succession during the experiments. Initially an Agilent 53230A counter in reciprocal continuous mode was used to count the beat frequency of a free-space double-pass AOM unit (300 MHz, results of Sec.~\ref{sec:clock_transfer_results}). After that, this setup was replaced by a zero-dead-time K+K FMX-50 counter to count the beat frequency of a fibre-coupled AOM (-42 MHz, results of Sec.~\ref{sec:roundtrip_instability_results}).

\begin{figure}[htb]
    \centering
    \includegraphics[width=\columnwidth]{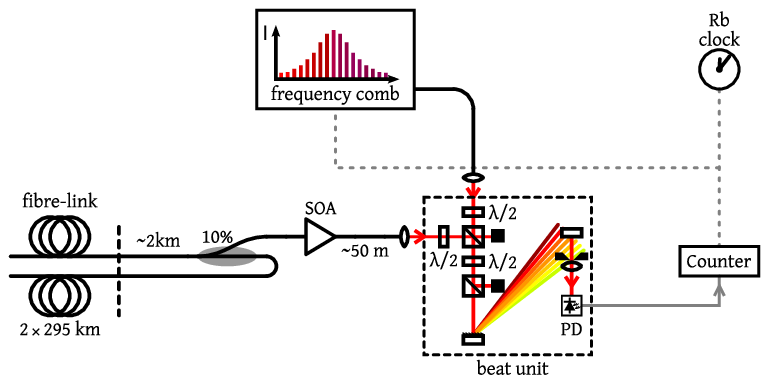}
    \caption{Experimental setup for the remote optical frequency measurement (Groningen). Of the received optical power 90\% is sent back to Amsterdam. To improve the signal of the free-space beat unit, the link light is amplified with a BOA-6434 semiconductor optical amplifier (SOA). The amplified light is then combined with light from the fibre frequency comb laser in a free-space beat unit to obtain an rf beat between the nearest frequency comb mode and the CW link laser.}\label{fig:fibre_network_experimental_setup_clock_comparison}
\end{figure}

The remote (Groningen) optical frequency characterisation setup is depicted in Fig.~\ref{fig:fibre_network_experimental_setup_clock_comparison}. The link laser is amplified using a semiconductor optical amplifier, and guided to a free-space beat unit. The beat frequency is counted using a K+K FMX-50 counter. The frequency comb laser and counters are frequency referenced to the Rb-standard.

\section{Measurements and simulations}\label{sec:measurement}
Two series of measurements have been performed on the fibre link, namely an 'optical-versus-GPS' measurement for which the optical frequency is measured simultaneously at Amsterdam and Groningen and compared, and a roundtrip measurement at the Amsterdam site (Fig.~\ref{fig:fibre_network_experimental_setup}). It is known that soil temperature variations of the fibre may lead to significant frequency instability
(see, for example,~\cite{lit:pap:ITIM-emardson-2008}). In this section, we present a model to describe and predict the influence of soil temperature variations on the frequency stability of underground fibre links. The results of the temperature model are compared with actual soil temperature and link stability measurements.

Apart from the instability contributed by the fibre link, the stability of the frequency transfer is limited by three sources. First, the stability of the frequency comb laser is limited by the Rb/GPS rf reference oscillator used to control the parameters of the comb. This reference has a specified relative instability of $<10^{-11}$ at 1 second and below $10^{-12}$ between $10^{3}$ and $10^{5}$ seconds, with a minimum around $3\pow{-13}$. This long-term frequency instability is transferred to the CW link laser through the various rf locks in the setup.

The rf locks themselves also contribute to the frequency instability, which we assess as follows. Using a second, similar, CW laser (RIO Orion) locked to the frequency comb, and employing the virtual beat note technique~\cite{lit:pap:APB-telle-2002}, the combined instability of the locking electronics is determined to be $<9.1\pow{-16}$ at 1 second. Thus, on time scales longer than 1s, the rf noise is effectively averaged out so that its influence on the link measurements may be neglected.

A second source of instability plays a role in the roundtrip measurements, for which a frequency drift of the laser source may lead to an apparent frequency shift of the (delayed) light transmitted by the fibre loop. Considering the 3.2 ms roundtrip propagation delay of the light over the fibre link, the (linear) drift of the frequency standard ($10^{-11}$ at 1 s) only plays a role at the level of $3.2\pow{-14}$ at 1 s, decreasing as $\sim1/\tau$, the inverse of the counter gate time, due to the fractional time overlap $\anbrace{\tau_{\text{gate}} - \tau_{\text{roundtrip}}}/\tau_{\text{gate}}$ between reference and roundtrip light in the frequency comparison. This is one order of magnitude smaller than the typically measured link stability at 1 s, and averages down more rapidly with increasing $\tau$.

For the optical-versus-GPS comparison, a third source of instability is due to the fact that the Rb clocks in Amsterdam and Groningen are independently locked to GPS time. Small intrinsic phase differences between GPS signals, received at geographically separated locations, thus propagate through the frequency locks, and may manifest itself as additional noise in the link frequency measurement.

\subsection{Link frequency transfer stability over \linklength}\label{sec:roundtrip_instability_results}
\begin{figure}[htb]
    \centering
    \includegraphics[width=\columnwidth]{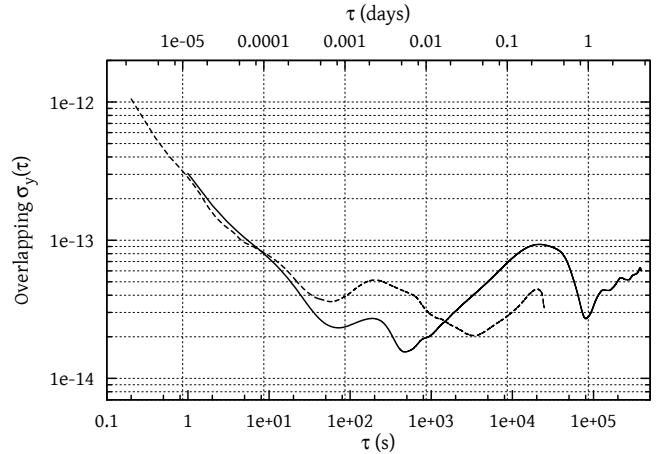}
    \caption{Comparison of two roundtrip stability measurements as overlapping Allan deviation (ODEV). (solid) Measurement of almost 9 days from 2013-09-30 to 2013-10-09 (outliers due to accidental low beat signal in this period where taken out and replaced with the median of the dataset, see text). The peak at 0.5 days and dip at 1 day are typical for frequency deviations with a one-day period. (dashed) Measurement of more than 13 hours performed at 2012-07-06, all data were included.}\label{fig:roundtrip_stability-120706_vs_130930}
\end{figure}

The intrinsic frequency transfer stability of a fibre link is best measured on a closed loop, so that the frequency of the input signal can be directly compared with that of the roundtrip signal. Figure~\ref{fig:roundtrip_stability-120706_vs_130930} shows the results of two roundtrip stability measurements. The first measurement is based on a 13-hour time series of frequency measurements acquired on 2012-07-06, using the free-space AOM unit (Fig.~\ref{fig:fibre_network_experimental_setup_roundtrip_measurement}) while recording with the Agilent 53230A. For this dataset the absolute frequency of the roundtrip optical signal was calculated and used to determine the overlapping Allan deviation (ODEV).

To compare the measurement with the soil temperature model, statistics on the scale of several days are needed. Therefore a longer roundtrip stability measurement took place from 2013-09-30 to 2013-10-09, using the fibre-coupled AOM while recording with the K+K FMX50 counter. This measurement contains a few periods during which the beat signal was too low for the FMX50 input circuits to record properly, leading to frequency outliers. Outliers were removed according to Chauvenets criterion ($P=0.5$, 4102 datapoints (0.52\% of the total set) removed) and were replaced with the median of the dataset. We have verified that the ODEV statistics are not influenced significantly by this operation.

\subsection{Clock transfer stability}\label{sec:clock_transfer_results}
\begin{figure}[htb]
    \centering
    \includegraphics[width=\columnwidth]{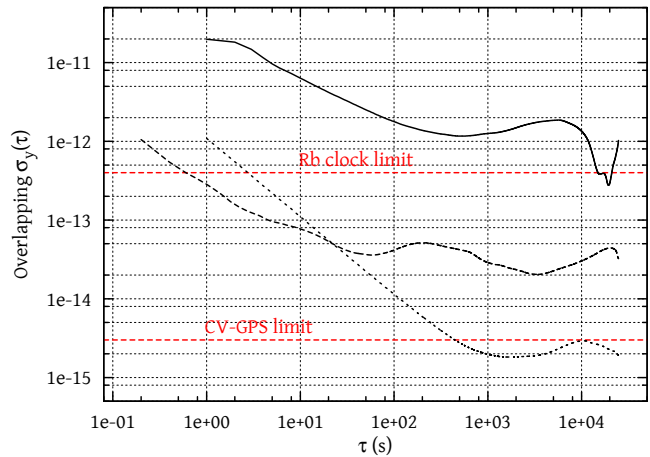}
    \caption{Comparison of ODEVs of the in-loop link laser stability relative to the frequency comb (short dash), the roundtrip stability (long dash), and the remote link laser frequency stability measured in Groningen (solid) of the 2012 measurement series. The (red) horizontal dashed lines indicate the Rb clock limit (SRS PRS10 datasheet), and the current best CV-GPS limit reported in~\cite{lit:proc:zhang-2009}.}\label{fig:link_measurement_120706}
\end{figure}

In the previous section we established that for averaging periods of more than 10 s the frequency transfer instability of the link is $<1\pow{-13}$. Given the stability of the laboratory frequency standards (Rb clocks) it is therefore to be expected that a direct measurement of the link laser frequency at the remote site (Groningen) yields the mutual clock stability. During the 2012 measurement session, the absolute laser frequency at the remote site was recorded. The result of this measurement, together with the roundtrip stability and the stability of the link laser lock frequency are presented in Fig.~\ref{fig:link_measurement_120706}.

The link laser frequency was determined with respect to the frequency comb, taking into account variations of \frep and \fceo. The determination of the absolute laser frequency is limited by the resolution of the \frep counter to $2.5\pow{-13}$/s. The maximum observable roundtrip instability is therefore about 1\pow{-15}/s. The observed instability is significantly higher, showing that the fibre may already influence the roundtrip stability at the second time scale.

The frequency of the laser used for the characterisation of the remote link was determined with respect to the nearest frequency comb mode. Together with the mode number determination, this directly delivers the remote absolute optical frequency. As can be seen in Fig. 7, the link laser frequency instability at the remote site is much larger than the total roundtrip instability. Therefore, the local frequency references are the limiting factor. This means that we have effectively performed a frequency comparison between the GPS referenced Rb atomic clocks over the fibre link. The measured performance is slightly worse than expected based on the Allan variance graph in the datasheet ($<10^{-12}$ at $>100$ s) of the clocks. This is probably due to imperfect GPS reception, and uncorrelated frequency noise of the fibre combs used in this experiment.

\subsection{Limits on frequency transfer stability due to soil temperature fluctuations}
Based on coarse estimates we expected (soil) temperature fluctuations to have a major influence on the passive stability of the fibre link. The Royal Netherlands Meteorological Institute (KNMI) provided us with soil temperature measurements taken at the Cabauw site (see Fig.~\ref{fig:surfnet_fibre_network}). Temperatures are measured at depths of 0, 2, 4, 6, 8, 12, 20, 30 and 50 cm  every 12 seconds and averaged over 10 minute intervals. Out of five KNMI locations the site at Cabauw is the only location for which soil temperature data with such high temporal resolution is available in the Netherlands. The provided datasets consist of one set covering the 9 day link measurement with a 12 second time interval in October 2013~\cite{lit:misc:KNMI-made_available_by}, and a 2 year dataset (2011, 2012) with a 10 minute interval~\cite{lit:web:KNMI-CESAR}.

Equation~\eqref{eq:frequency_from_temperature} is used to convert these temperature series to frequency deviations expected on a \linklength{} fibre link with the given thermal expansion and thermo-optic coefficients. In the next sections we use these reference data to compare the link stability with the soil temperature model, Eq.~\eqref{eq:soil_temperature_hoeven}.

\subsubsection{Soil temperature and frequency transfer stability}
The raw frequency data from the 9-day link stability measurement is compared to the frequency deviations as derived from the KNMI soil temperature dataset over the same time period. The first question that needs to be addressed is to what extent such a comparison, correlating temperature effects measured in locations separated by tens of kilometres, is meaningful. The Cabauw measurement site (Fig. 1) is located more than 40 km south of the (geographically) 200 km long trajectory of the fibre. Therefore the exact trends in soil temperature at the Cabauw site and along the fibre link can differ, due to local variations in solar irradiation and precipitation. Nevertheless, significant temperature correlations are expected as the different locations are relatively close, as seen from a meteorological and climatological point of view.

\begin{figure}[htb]
    \centering
    \includegraphics[width=\columnwidth]{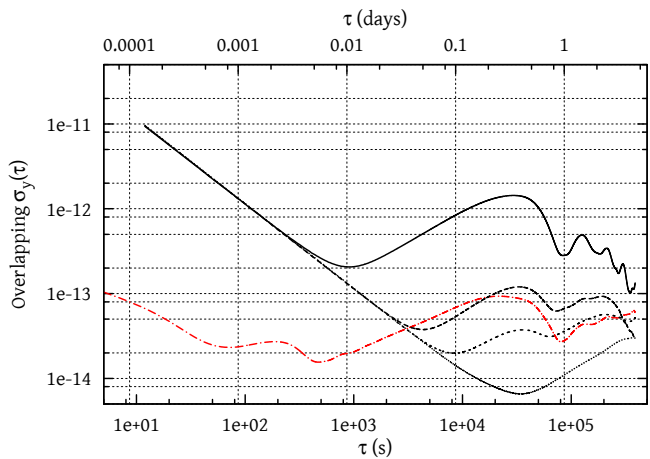}
    \caption{Fibre-link roundtrip stability (dash-dot) compared with roundtrip stabilities as calculated from the KNMI soil temperature measurements for different fibre depths: on the surface (solid curve), at 20 cm depth (long-dashed curve), at 30 cm depth (short-dashed curve), and at 50 cm depth (dotted curve). At shorter averaging times, the model curves display a $1/\tau$ slope, which indicates that on shorter time scales temperature noise is significantly more prominent in the KNMI measurements than in the temperature-dependent link stability.}\label{fig:roundtrip_comparison_measurement_knmi}
\end{figure}

Figure~\ref{fig:roundtrip_comparison_measurement_knmi} shows the roundtrip stability in comparison to the fibre-link stability calculated from the KNMI soil temperature data. KNMI aims at a relative accuracy among sensors of 0.01 K and an absolute accuracy of 0.1 K. The noise in the soil temperature measurements amounts to a few mK.  More accurate soil temperature observations will be difficult to make, let alone capturing all relevant variations along the optical path. All stabilities derived from soil temperature show a $1/\tau$ slope for shorter time scales, which indicates the presence of white noise in the temperature measurements that does not appear in the fibre link instability. The noise levels at these time scales obscure frequency drift due to the daily temperature cycle. This cycle leads to the rising slope at longer time scales, with a local maximum at an averaging time of half a day. On even longer time scales, this instability cycle continues with a one-day period, and with local maxima decreasing in height over time. This is most clearly seen in the surface temperature curve.

The measured roundtrip stability curve shows averaging at short time scales less than 100 s, while at time scales larger than 1000 s frequency drifts due to the diurnal temperature cycle dominate the instability. The origin of the level of instability at time scales less than 4\pow{3} s remains unclear. Possible causes are the several hundreds of meters of the fibre link located inside buildings, which are subject to significant and relatively fast temperature variations (due to e.g. airconditioning systems), while other factors like stress induced frequency fluctuations can not be entirely ruled out. For averaging times larger than 4\pow{3} s, the link measurement shows qualitative agreement with the KNMI data at depths of 20--30 cm.

\begin{figure}[htb]
    \centering
    \includegraphics[width=\columnwidth]{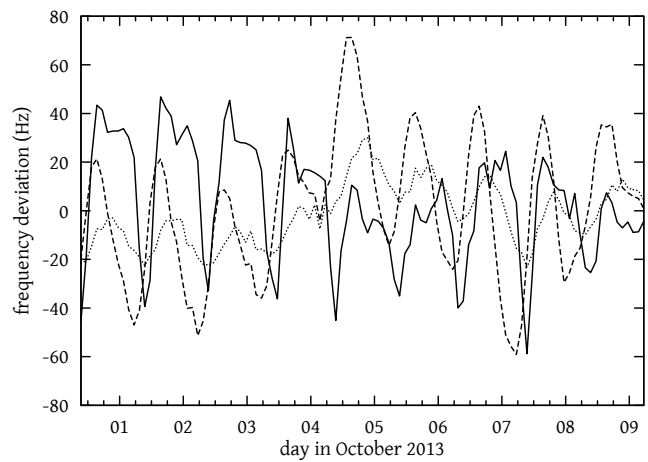}
    \caption{Frequency deviations after a roundtrip through the fibre link (solid) and frequency deviations calculated from the soil temperature data at 20 cm (dashed) and 30 cm depth (dotted).}\label{fig:roundtrip_time_series}
\end{figure}

The correlation between soil temperature and link frequency drift can be inferred more directly from the temperature and frequency measurement time series. To this end, the roundtrip frequency deviations from the AOM frequency, $\Delta f$, are compared to the frequency deviations $\Delta f_{n}$ due to soil temperature $T_{\text{KNMI},n}$ at depth $n$, estimated using Eq.~\eqref{eq:frequency_from_temperature}. Both datasets are averaged over two-hour windows to reduce noise levels. Figure~\ref{fig:roundtrip_time_series} shows the raw data of the nine-day link measurement in comparison with $\Delta f_{n}$ at 20 cm and 30 cm depth (2 hour averages).

The depth at which the fibre link is buried is not precisely known, and it furthermore may vary along the path of the link. An estimate of the 'effective' depth of the link may be obtained by assigning a weighting factor $c_{n}\ge0$ to each time series $\Delta f_{n}$, and minimising (by least squares fitting) the difference $y$ between the measured and calculated frequency deviations
\begin{equation}
    y=\Delta f - \sum_{n=0}^{N} c_{n}\Delta f_{n}(T_{\text{KNMI},n}) + f_{\text{offset}}.
\end{equation}
The frequency offset parameter $f_{\text{offset}}$ is needed to include a systematic offset between the link measurement data and the KNMI data. Such an offset may be caused by an overall relative temperature change between the fibre link path and the Cabauw site.

\begin{figure}[htb]
    \centering
    \includegraphics[width=\columnwidth]{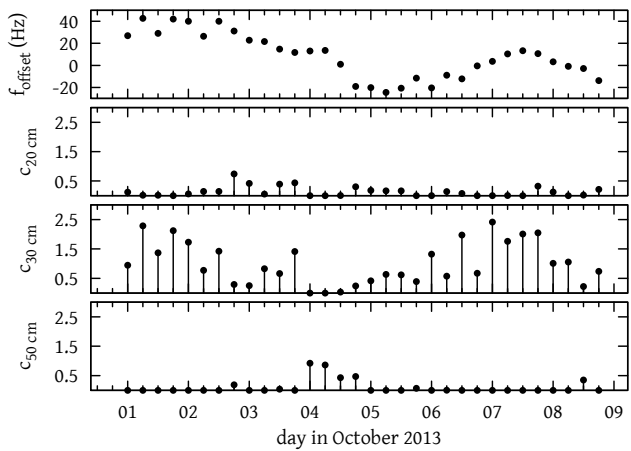}
    \caption{Fit parameters obtained by least-squares fitting to (partly overlapping) 24-hour subsets of roundtrip frequency data, with a spacing of six hours between each subset. Day of year represents the centre of the data range. (top) frequency offset for fit. (bottom three panels) Values of the $c_{n}$ for the most important depths; the $c_{n}$ found for the other depths are negligibly small. Averages and standard deviations over this dataset are $f_{\text{offset}}=8.1(20.3)$ $c_{\text{20 cm}}=0.13(0.17)$, $c_{\text{30 cm}}=1.01(0.72)$, and $c_{\text{50 cm}}=0.10(0.24)$.}\label{fig:roundtrip_time_fit_parameters}
\end{figure}

Figure~\ref{fig:roundtrip_time_fit_parameters} shows the variation of the fit parameters $c_{n}$ and $f_{\text{offset}}$ over time. These parameters and their time dependence are obtained as follows. First, all input data (i.e. the measured frequency deviations and the frequency deviations estimated from soil temperature measurements) are averaged with a one-hour window. Of the averaged data sets, a 24-hour subset is taken (labeled by the median of the time stamps in the set), for which the coefficients $c_{n}$ and $f_{\text{offset}}$ are found by least-squares fitting. This last step is repeated for a 24-hour subset which is offset by six hours with respect to the previous subset, until the entire data set is covered. From Fig.~\ref{fig:roundtrip_time_fit_parameters} it follows that best agreement is found for an average fibre depth of about 30 cm. The least-squares fit method yields solutions which are generally well aligned in phase with the measured data (Fig.~\ref{fig:roundtrip_time_series}). However, Fig.~\ref{fig:roundtrip_time_series} also shows that the agreement between the amplitude of the frequency deviations at 30 cm depth and the measured data is poor.

In the time series of the link measurement (Fig.~\ref{fig:roundtrip_time_series}) the frequency deviations before 3 October and after 6 October appear to follow the predictions based on KNMI data, while between 3 October and 6 October the curves seem to be substantially less correlated. This discrepancy might be linked to the fact that during this period, the fibre-link path received considerably more precipitation on 3 October 2013 than the measurement site at Cabauw~\cite{lit:web:KNMI-KDC}. Moreover, solar irradiation differed substantially from day to day between Cabauw and the fibre link in this period~\cite{lit:web:KNMI-KDC}. It is conceivable that this led to local soil temperature variations along the fibre link and, thus, to the observed discrepancy. This behaviour illustrates the limited power of the temperature model for predicting or estimating instantaneous frequency variations based on soil temperature measurements.

\subsubsection{A soil temperature model for frequency transfer stability estimation}
Soil temperature data can be used to predict the long-term stability of fibre links, but it can be burdensome to obtain or construct long historical records. For example in 1961 soil temperature was measured 3 times a day~\cite{lit:rep:woudenberg-1966}. Also, soil temperature is less often measured at meteorological measurement sites than other quantities. In the Netherlands KNMI has such data available for only four sites. In case of scarce soil temperature data, the sinusoidal soil temperature model (Eq.~\eqref{eq:soil_temperature_hoeven}) can be used to construct an artificial temperature cycle by superposition of a diurnal and an annual temperature cycle, which can be used to estimate the frequency transfer stability of fibre links. Such models can also be used to predict frequency transfer stability for various types of soil.

The amplitude of the diurnal temperature variation itself varies approximately sinusoidally during the year, and is given by
\begin{equation}
    A_{T_{\td,\text{year}}}(t)=T_{\td,\text{year}} + A_{\td,\text{year}}\sin\anbrace{\frac{2\pi}{P_{\text{year}}}(t-t_{0,\text{year}})},
\end{equation}
where $T_{\td,\text{year}}$ is the average diurnal temperature variation, $A_{\td,\text{year}}$ is the amplitude of the annual variation of the diurnal amplitude, $P_{\text{year}}$ is the annual period, and $t_{0,\text{year}}$ is used to shift the temperature cycle to fit the model to the long term measurement data of KNMI.

The total annual temperature cycle now becomes
\begin{equation}\label{eq:yearly_soil_temperature_cycle}
    T_{\text{annual}}(z,t) = T_{0} + T_{\text{day}}(z,t,A_{T_{\td,\text{year}}}(t)) + T_{\text{year}}(z,t),
\end{equation}
where $T_{\text{day}}$ and $T_{\text{year}}$ are given by Eq.~\eqref{eq:soil_temperature_hoeven}, but now with the amplitude $A_{T_{0}}$ of the diurnal variation being a function of time.

We obtain a set of model parameters for Eq.~\eqref{eq:yearly_soil_temperature_cycle} by fitting Eq.~\eqref{eq:soil_temperature_hoeven} to the data from Van der \nobreak{Hoeven} and \nobreak{Lablans}~\cite{lit:rep:hoeven-1992} as an estimate of the annual variations, and to the data from \nobreak{Woudenberg}~\cite{lit:rep:woudenberg-1966} to estimate the cycle of diurnal temperature variations. The soil constant was taken $C_{s}=7.5\pow{-4}$, which is representative for sand (being in a state between wet and dry), relatively dry loam, and clay~\cite{lit:rep:hoeven-1992}. The obtained average, amplitude and phase values are given in Table~\ref{tab:soil_model_fitted_parameters}.

\begin{table}[htb]
    \tiny
    \centering
    \begin{tabular}{ l | c c c }
        \hline
        & Offset (\degrees C) & Amplitude (\degrees C) & $t_{0}$ (s) \\
        \hline
        Annual variation (surface) & 10.2 & 8.8 & 9.64\pow{6} \\
        Annual day amplitude $A_{T_{\td,\text{year}}}$ &  2.3 & 1.4 & 7.94\pow{6} \\
        Diurnal variation &  0.0 & $A_{T_{\td,\text{year}}}(t)$ & 3.67\pow{4} \\
        \hline
    \end{tabular}
    \caption{The parameters of the soil temperature model of Eq.~\eqref{eq:yearly_soil_temperature_cycle} retrieved by a least-squares fit to data obtained from~\cite{lit:rep:hoeven-1992} and~\cite{lit:rep:woudenberg-1966}.}\label{tab:soil_model_fitted_parameters}
\end{table}

\begin{figure}[htb]
    \centering
    \includegraphics{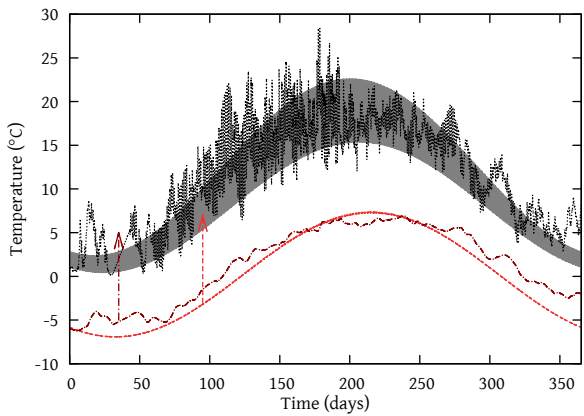}
    \caption{Yearly variation of soil temperature at various depths. (solid, appearing as a wide band due to diurnal variations which are not resolved at the time scale of the plot) Modelled temperature at the surface, (dashed) at 50 cm depth. (dotted) Measured temperature at the surface, (dash-dotted) 50 cm depth. Temperatures at 50 cm depth are lowered by 10 degrees centigrade for visibility, arrows indicate true position.}\label{fig:soil_temperature_yearly_fluctuation}
\end{figure}

The soil temperatures resulting from the model are plotted in Fig.~\ref{fig:soil_temperature_yearly_fluctuation} and compared with the measured KNMI data set covering the year 2011. The plot compares the data at the surface and at 50 cm depth and shows that the model is indeed in reasonable agreement with direct soil temperature measurements.

\begin{figure}[htb]
    \centering
    \includegraphics{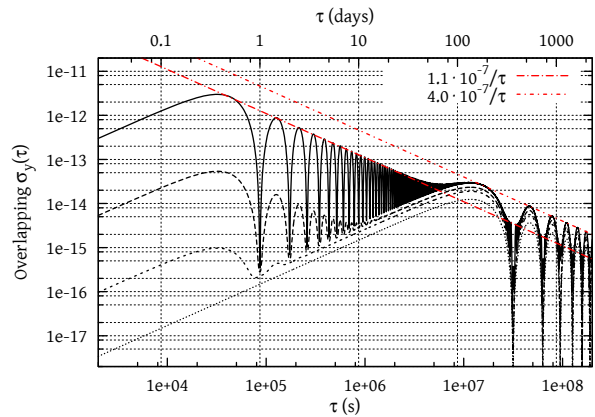}
    \caption{Calculated link roundtrip stability resulting from soil temperature variations according to the model (Eq.~\ref{eq:yearly_soil_temperature_cycle}) at several depths. Frequency stability at (solid) the surface, (long dashed) 50 cm depth, (short dashed) 100 cm depth and (dotted) 200 cm depth. The dash-dotted lines indicate the $1/\tau$ behaviour with a maximum instability of $2.6\pow{-12}$ at half a day for the diurnal variation and $2.5\pow{-14}$ at half a year for the annual variation.}\label{fig:soil_temperature_allan_deviations}
\end{figure}

A long time series of soil temperatures was generated from the model, and estimates of the long term frequency stability of a hypothetical fibre link of the Amsterdam -- Groningen type were calculated using Eq.~\eqref{eq:frequency_from_temperature}. Figure~\ref{fig:soil_temperature_allan_deviations} shows the calculated stabilities at the day-to-year time scale, and for various depths. From the graph it is clear that more deeply buried fibres offer a substantial stability improvement at the timescale of days, while the stability at the year scale is much closer to that of fibre buried closer to the surface.

\begin{figure}[htb]
    \centering
    \includegraphics{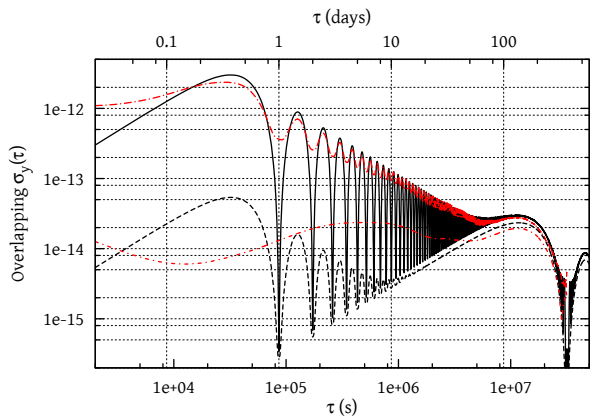}
    \caption{Frequency stability comparison between soil temperature measurement and model. (solid) Model, surface. (long-dash dotted) Measurement surface. (dashed) Model 50 cm depth. (short-dash dotted) Measurement 50 cm depth.}\label{fig:soil_temperature_model_measurement}
\end{figure}

Figure~\ref{fig:soil_temperature_model_measurement} compares the frequency stability, computed using Eq.~\eqref{eq:frequency_from_temperature} and the soil temperature measurement series of 2011 and 2012, with that obtained from Eq.~\eqref{eq:frequency_from_temperature} and the sinusoidal model, Eq.~\eqref{eq:yearly_soil_temperature_cycle}. The most prominent feature is the discrepancy for 50 cm depth, likely due to fluctuations in weather conditions on long (days/weeks/months) time scales, which cause a higher instability at time scales between a day and half a year. These fluctuations are clearly visible in Fig.~\ref{fig:soil_temperature_yearly_fluctuation}. This noise also leads to a smoothing of the strong minima in the frequency instability curves, which are a consequence of the sinusoidal temperature behaviour.

\section{Conclusions and outlook}\label{sec:conclusions}
We investigated the passive frequency stability of a \linklength{} carrier-grade, unidirectional fibre link between VU University Amsterdam LaserLaB and Van Swinderen Institute, University of Groningen, using a single DWDM channel, and with live optical data traffic present in other DWDM channels. The observed frequency instability of the link lies in the range $10^{-14}$ to $10^{-13}$ for averaging times 10 -- 8\pow{5} s. This result implies that such fibre links are well suited to distribute the frequency of commercial Rb atomic clocks with negligible loss of accuracy. A model for thermo-optical fibre length variation was developed which relates frequency variations to soil temperatures as a function of depth. We employed this model taking actual long-term soil temperature measurements as input, as well as predictions obtained from existing soil temperature models. We combined link frequency measurements with soil temperature measurements at the KNMI Cabauw site to show that the observed link stability corresponds to an average fibre depth of about 30 cm.

Qualitative agreement is found between the soil temperature model and KNMI measurements, while predictions of frequency stability based upon this model agree with actual roundtrip frequency measurements to within an order of magnitude. Although the predictive power of the soil-temperature model is limited, it does provide insight into the relation between soil temperature, fibre-optic path length variations and frequency-transfer stability. Our model thus allows estimating the passive frequency stability of fibre links for averaging times ranging from days to years, and allows to estimate upper bounds on the passive link instability. Such information will be useful for the design of future one-way frequency distribution systems based on underground fibre-optic infrastructure.

The presented results show that soil temperature fluctuations have a large impact on the passive frequency stability of optical carriers over underground fibre links for time scales longer than approximately 1000 seconds. The results of our study confirm the conclusions of previous work that fibre-optic infrastructure is sufficiently stable for one-way atomic clock frequency distribution over hundreds of kilometres distance, and with $1\pow{-13}$ relative instability~\cite{lit:pap:JOSAB-williams-2008,lit:proc:jaldehag-2009}. This figure compares favourably to the stability of commercial GPS-disciplined Rb clocks. The residual frequency variations are sufficiently small to back up the local oscillator of GPS-referenced clocks with indefinite holdover (provided a non-GPS-referenced master clock is used). This opens up the perspective of a terrestrial ``flywheel'' oscillator, embedded in the fibre-optic telecommunications network, which can be used to back-up GPS-referenced oscillators during periods of GPS outage.

\section*{Acknowledgement}
This work was supported by the Netherlands Foundation for Fundamental Research of Matter (FOM) through the program ``Broken Mirrors \& Drifting Constants''. J.C.J.K. thanks the Dutch Organisation for Scientific Research (NWO) and the Dutch Technology Foundation (STW) for support.

\end{document}